\documentclass[twocolumn,showpacs,preprintnumbers,amsmath,amssymb]{revtex4}

\usepackage{graphicx}
\usepackage{dcolumn}
\usepackage{bm}

\begin{document}

\preprint{}
 
\title{Non Trivial Effects in Heavy Nuclear Systems\\due to Charge 
Rearrangement}

\author{B. Toledo}
\email{btoledo@fisica.ciencias.uchile.cl}
\author{J. Rogan}
\author{C. Tenreiro}
\author{J. A. Valdivia}
\affiliation{
Universidad de Chile, Facultad de Ciencias, Departamento de F{\'\i}sica
}

\date{\today}

\begin{abstract}
In this report we propose that a charge rearrangement within heavy
ions may occur during fusion reactions, lowering the Coulomb barrier as the nuclear systems approach each other. In this situation a modified Gamow factor has been calculated, which emphasizes the importance of spatial degrees of freedom in nuclear matter.
\end{abstract}


\maketitle

\section{Introduction}

It is commonly believed that the point-like approximation for the
Coulomb potential is accurate enough when describing the nuclear
dynamics during a fusion interaction \cite{vn}\cite{ah}. In this
context, there are a number of anomalies observed in fusion reactions
near the Coulomb barrier \cite{ct}\cite{h}, and a number of anomalies
in the abundances of \emph{p-elements} produced in supernovas \cite{cr}, specially the production of rare nuclide, that do not fit well in our present understanding of supernova and nuclear dynamics. 

In this work, we will consider the effect of a spatial redistribution
of charge during a nuclear interaction, which may add an extra
ingredient when considering the anomalies observed in fusion reactions
near the Coulomb barrier \cite{ct}\cite{h}. The key idea in this model
is to take each nuclei as a \emph{plasma ball} (inspired in a
\emph{quark-gluon plasma} \cite{bm}, although no calculation presented
here has any relation to that specific subject), that has the ability
to redistribute some of the charge it contains in the presence of the field produced by the second nuclei. As long as distances between the bodies are large enough, we can disregard the strong interaction between them, and consider only the coulomb interaction. It is important to notice that in this context the strong force plays a role within each body, making the transport of charge within the nuclei possible. Even though the dynamics of our model is developed in a quantum mechanical context, it does not consider spin effects, nor other non classical effects apart from those described by the scalar Sch{\"o}dinger equation. 

Our main result is that the redistribution of charge inside each ball
lowers the Coulomb barrier while ions approach each other. The
modification is expressed through the corrected differential cross sections and the Gamow factor, which is of special interest in astrophysical scenarios (e.g., reaction networks in supernova \cite{sw}). 

\section{The Model}

Charge rearrangement in the presence of an external electric potential
is common in nature. For example, charge move within conductors in
response to external electric fields, as described by distance dependent image charges.

In the case of nuclear reactions, there are a number of ways in which we can expect to obtain a charge rearrangement within a nuclei, in the presence of the field produced by a second nuclei.  For example we can imagine (1) a collective excitation of the nuclei, that is a fluctuation of nuclear matter around its equilibrium state; (2) a fluctuation involving only the surface, with an unperturbed bulk, of a proton rich nuclei; (3) a local interaction affecting an unpaired proton or neutron; (4) etc. In all of these cases the specific description will be strongly dependent on the system under consideration. 

For this paper, and simplicity, lets assume that this charge
rearrangement inside the nuclei can be modeled, in a first
approximation, as an induced electric dipole. More exact calculations
can also be considered, such as higher order moments, but for
simplicity will be reported elsewhere. We may imagine many others ways
to induce an electric dipole moment in a nuclei, all of which have
their own lifetime. Of interest is the case that involves
time-reversal violation \cite{fz}, but produces an effect that seems
to be weaker than those coming from charge rearrangement \cite{rg}. It
is important to note, that the rearrangement of charge will depend on
the position and charge of the second nuclei. 

For the problem at hand, we can write the electric dipole operator as
\begin{eqnarray}\label{do}
\mathbf{D}=e\,\frac{N}{A}\sum_{p=1} ^Z\mathbf{x}_p-e\,\frac{Z}{A}\sum_{n=1} ^N \mathbf{x}_n\,,
\end{eqnarray}
where $e$ is the unit charge, $A$ is the mass number, $Z$ is the
atomic number, and $N$ the neutron number. The operator $\mathbf{D}$
has a well defined expectation value during the lifetime of the charge rearrangement, $\mathbf{p}=\langle \mathbf{D}\rangle$. Since the strength of any multipolar field is proportional to the size of the polarized object \cite{ja}, we will consider interactions of type $A+p$, where $A$ is a large ion and $p$ is a proton. We will neglect a possible dipole moment for the proton.

Suppose now that the incoming proton induce a charge rearrangement in the ion, observed externally as a dipole moment, which can be visualized as a pair of charges $\pm q_e$ displaced a distance $d$ away from the ion center. The orientation of the dipole is induced in such a way as to reduce the repulsion interaction, for a given separation $r$ between the ion and the proton. Therefore, in the  $\lim_{d\to 0} 2 q_e d=p_z$, we can approximate the potential energy as

\begin{eqnarray}
U(r)=\frac{Z_1\,Z_2\,e^2}{r}-\frac{p_z Z_2\,e}{r^2}\,,
\end{eqnarray}
where $p_z\approx |\mathbf{p}|$, and the minus sign comes from the discussion above. We are interested in the potential close to the origin, and in the limit $d\ll r$. Since, the rearrangement of charge, as described by $p_z$, will depend on the relative position of the second nuclei, we are interested in situations for which $p_z$ is finite as $r\to0$. This definition is consistent with our above assumptions.

Due to the symmetry of the problem we choose parabolic coordinates: $\xi=r-z$, $\eta=r+z$, \mbox{$\varphi=\arctan(y/x)$} \cite{rn}, and seek a solution of the
form

\begin{eqnarray}
\psi=e^{ikz}\phi(\xi)\,,
\end{eqnarray}
where the function $\phi$ corresponds to a spherical outgoing wave. Operating on the scalar Schr{\"o}dinger equation, we obtain

\begin{eqnarray}\label{imecucomp}
&&\xi \,\phi ''(\xi )+ \left( 1 - i \,k\,\xi  \right) \,\phi '(\xi )-\left(\kappa
  k-\frac{\beta^2}{\xi}\right)\,\phi(\xi)=0\ ,\qquad\\
&&\kappa=\frac{Z_1Z_2\alpha c}{v},\,\quad\beta=\frac{2}{\hbar}\sqrt{\frac{\mu p_zZ_2 e}{\lambda}},\quad \lambda=\frac{1}{1-\cos \theta},\nonumber
\end{eqnarray}
where $k$ is the wave number, $\mu$ the reduced mass, $c$ is the speed of light, $v$ is the speed of the projectile, $\alpha$ is the fine structure constant, and $\theta$ is the azimuthal outgoing angle for the projectile (this must be small to satisfy $p_z\approx |\mathbf{p}|$). This angle has meaning only for an elastic interaction.

This equation has two linearly independent solutions that are regular at the
origin, one for each possible sign of $\beta$, so the most general solution is a linear combination of both

\begin{equation}
\label{imsolcomp}
\begin{split}
\phi(\xi)&=C_+\,\xi ^{i\beta} {}_1 F_1(-i\,\kappa+ i\beta,1+ 2\,i\,\beta,i\,k\,\xi)\\
& \quad +\,C_-\,\xi ^{-i\beta} {}_1 F_1(-i\,\kappa- i\beta,1-2\,i\,\beta,i\,k\,\xi)\ ,
\end{split}
\end{equation}

\noindent where $_1 F_1$ is the hypergeometric confluent function, and $C_{\pm}$ are the two constants of integration. Now we want to calculate the Gamow factor, which is by definition the quotient between the probabilities of finding close together particles with the same and opposite sign \cite{rn}. In order to obtain a value for the constants, we normalize the flux asymptotically far and make \mbox{$C_+=C_-$} for simplicity. In this way we can define the following function

\begin{equation*}
\label{rho}
\begin{split}
\rho(\kappa,\beta)&=\Bigg[\frac{2\beta\,\sinh\, \pi(\kappa+\beta)}{(\kappa+\beta)\,\sinh\, 2\pi\beta}\,e^{-\beta\pi}
+\frac{2\beta\,\sinh\, \pi(\kappa-\beta)}{(\kappa-\beta)\,\sinh\, 2\pi\beta}\,e^{\beta\pi} \\
&\hspace{-0.3cm}+\frac{4\beta}{\sinh\,2\pi\beta}\sqrt{\frac{\sinh\, \pi(\kappa-\beta)\,\,\sinh\,
    \pi(\kappa+\beta)}{\kappa^2-\beta^2}}\,\cos \vartheta_{\beta}\Bigg]\ ,\\
\vartheta_{\beta}&=\textrm{arg}\left[\frac{\Gamma(1+2i\,\beta)}{\Gamma(1-i\kappa+i\beta)}\,\frac{\Gamma(1+2i\,\beta)}{\Gamma(1+i\kappa+i\beta)}\right]\ ,
\end{split}
\end{equation*}

\noindent 
where the function \emph{arg} is the argument of the complex number ($z=re^{i\theta}$, arg $z=\theta$). From this expression we can write the corrected Gamow factor 

\begin{eqnarray}\label{gamowcor}
G_f=\left|\frac{\rho(\kappa_-,i\beta)}{\rho(\kappa_+,\beta)}\right|\ ,
\end{eqnarray}

\noindent where $\kappa_{\pm}\gtrless 0$ and $|\kappa_{\pm}|\gg 1$. Note that change
from $\beta\to i\beta$ due to the change in the orientation of the
dipole. To check this expression we take the limit $\beta \to 0$, corresponding to no rearrangement of charges, and we get

\begin{eqnarray}
\lim_{\beta \to 0} G_f=e^{-2\,\pi\,\frac{Z_1\,Z_2 e^2}{\hbar\,v}}\ ,
\end{eqnarray}

\noindent that is, we recover the usual expression. In this way we have calculated a corrected Gamow factor due to charge rearrangement within the target nuclei.

\section{analysis and results}

\begin{center}
\begin{figure}[!h]
\includegraphics[width=0.48\textwidth]{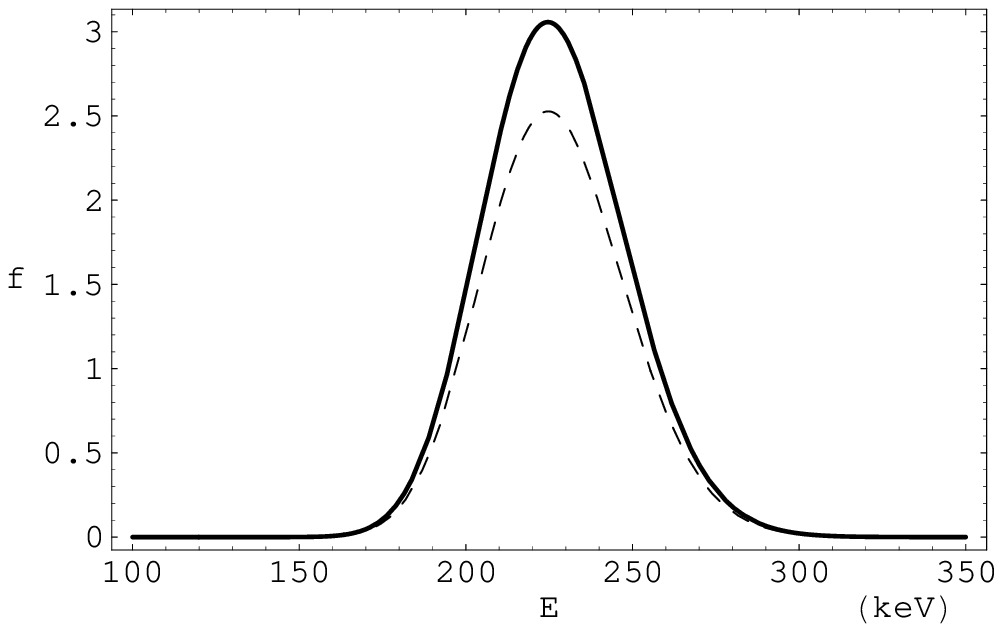}\\
\includegraphics[width=0.48\textwidth]{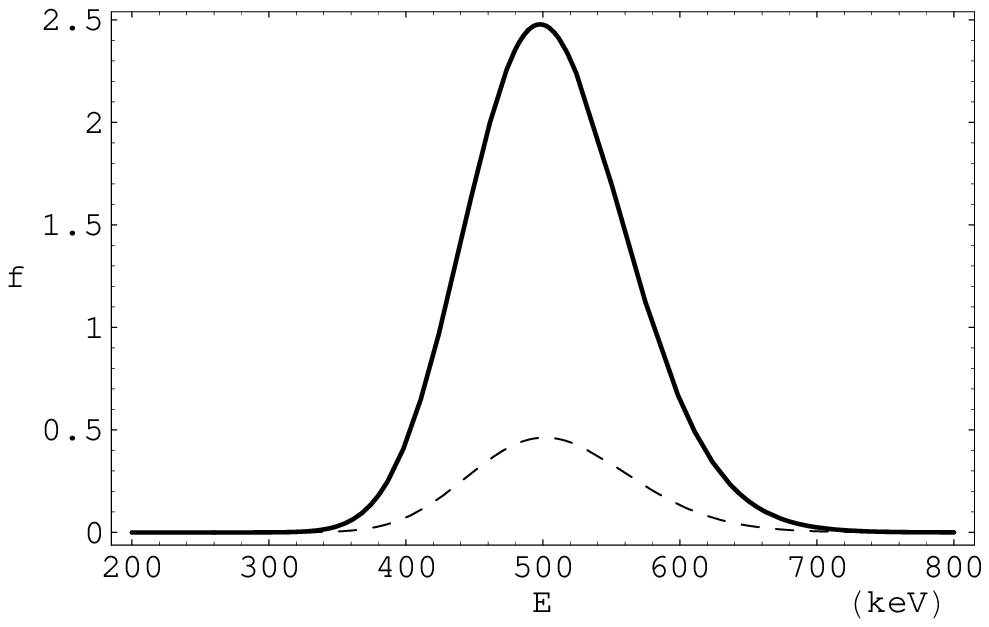}
\caption{\textbf{Gamow window:} The dashed line corresponds to the non corrected case and the solid line represents the corrected case for $p_z=0.122$($e$-fm) and $\theta=0.00012$(rad).
We use in this example the interaction ${}^{179}$Hf+$p$, for 
$kT=3$(keV). At the bottom the same reaction but for $\theta=0.00039$(rad) and $kT=10$(keV). (f$\,=g_{_f}(E,\beta)\,e^{-\frac{E}{kT}}$, where $g_f$ is $G_f$ 
as a function of $E$ instead of $\kappa$, and have been scaled by $10^{98}$ and $10^{65}$ respectively).}
\label{f1}
\end{figure}
\end{center}

With the aid of this simple model we can estimate in a first
approximation how the rearrangement of charge in ions may modify the
fusion process. As an example consider Fig. (\ref{f1}) in which we
plot the Gamow window that is of interest in astrophysics processes 
\cite{trs}. Here we make an estimation for a reaction
${179}\atop{72}$Hf+$p$. In particular, this heavy ion has an E1
transition between levels with $J^{\pi}$: 7/2$^-$ and 9/2$^+$, energy
E$_{\gamma}$=214.3(keV), and half life 1.86(ns) \cite{ion}. We take the ion
in this excited state, and in order to make a simple and rough estimation, we consider a semi-classical calculation starting from eq.(\ref{do}), which can be re-casted in the form
\begin{eqnarray}\label{dip}
p_z=e\frac{N\,Z}{A}\,d
\end{eqnarray}
where $d$ is the distance between the center of mass for proton and
neutron in the heavy ion. Now we take the energy associated to this
separation $d$ equal to E$_{\gamma}$ as the work done in a Yukawa type
potential \cite{lmk}. This estimation give us $p_z=0.122$(e-fm), which
is not too far from experimental measurements in similar ions
\cite{smk}. Now we need to estimate the azimuthal outgoing angle $\theta$,
and in the same spirit, we use the Rutherford relation between impact parameter $s$ and $\theta$ \cite{gol},
\begin{eqnarray}\label{dip1}
\theta=\pi-2\cot^{-1} \left(\frac{2Es}{Z_1Z_2 e^2}\right)
\end{eqnarray}
where $E$ is the energy of the incident particle. Taking $E=kT$, and
for $s$ the diameter of the proton in order to maintain $\theta$ small, we
have that for $kT=3$(keV) we obtain $\theta=0.00012$(rad), while for
$kT=10$(keV) we obtain $\theta=0.00039$(rad). As can be seen in the figure, the impact of this factor is strongly dependent not only on the specific system considered but also on thermodynamic variables.

Reaction cross section, which includes Gamow factor through the
astrophysical factor, is proportional to the amplitude of this
curve. Therefore, the rearrangement of charge could be important in
reaction networks in stellar interiors, permitting the burning process
of heavy elements to start with less energy than in the case in which there is no charge rearrangement.

At this point, is important to notice that this charge rearrangement effect is being considered in a general framework, and any detail in subsequent products will depend on the specific microscopic nuclear model used. For example, if we choose the Hauser-Feshbach formalism, the mechanism here proposed only affects the pre-compound nucleus stage, changing the number of such systems for the same incident energy but no its characteristics, leaving any subsequent nuclear evolution qualitatively unaltered.

Within the present model, the charge fraction that participates in the
charge rearrangement, and hence the dipole moment $p_z$, can not be
derived and we need to estimate it from other considerations
(parameters $p_z$ and $\theta$ given in the figures should be considered for example purposes only). We are developing a technique similar to molecular dynamic simulations (nucleon clusters) which may permit us not only estimate the charge fraction that participates in the charge rearrangement, but also it may provide us with the shape of the observed heavy ions as explained through charge rearrangement. The inclusion of the spin effects, which have not been considered in this analysis, may become relevant when taking into account selection rules due to angular momentum conservation. These considerations will be presented elsewhere.

Now, a comment on the approximation $p_z\approx |\mathbf{p}|$ is necessary. In eq. (\ref{do}), the position operators $\mathbf{x}_k$ are subject to fluctuations \cite{eb} as reflected in the vector dipole moment, therefore, the dipole vector is not always pointing in the exact direction of the incoming projectile. This allows for the possibility of a finite $\lambda$, implying an increase in the Gamow factor. On the other hand, it is possible that there are other contributions to the dipole moment, besides the induction effect. For example the nuclei could absorb electromagnetic radiation before the encounter with the proton, generating a dipole vector that will tend to align to the field of the incoming particle, but not exactly because of the fluctuations. Note that the requirement of a small $\theta \neq 0$ is compatible with the assumptions made in the calculation of the scattering cross section for the Coulomb field \cite{rn}.

In the scope of nuclear astrophysics, the electric dipole strength in
the energy region close to the neutron threshold has shown a strong
influence on various processes of nucleosynthesis because in relevant
scenarios even sub percent contributions to the E1-energy weighted sum
rule in this region may lead to completely different abundance
patterns \cite{go} \cite{hz}. Of course, this has an strong connection
to p-process \cite{hh}. 

It is not too far fetched that the proposed idea may also contribute
in the longstanding puzzle observed in laboratory measurements, in which the electron screening effect is surprisingly larger than theoretical prediction based on an atomic physics model \cite{ha2}.

\section{conclusion}

We have found that approximating the source of the coulomb field by a
point may not always provide an accurate description in a fusion
interaction, because it hides spatial degrees of freedom, and may be
one source of disagreement between the theory and experimental
measurements. The rearrangement of charge in ions may tend to increase
the reaction cross section, and could provide an extra ingredient to
the understanding of some anomalies in reactions near the coulomb barrier without modification of the nuclear potential. 

\section{ACKNOWLEDGMENTS}
This project has been financially supported by FONDECyT under contracts 
N$^{\circ}$1000676, N$^{\circ}$8990005, N$^{\circ}$1010988, 
N$^{\circ}$1030957, N$^{\circ}$1030727, one of us (B.T.) also acknowledges 
CONYCyT for a doctoral fellowship.

\end{document}